# The Sound Emission Board of the KM3NeT Acoustic Positioning System

**C.D. Llorens**[*], **M. Ardid**[♣], **T. Sogorb, M. Bou–Cabo, J.A. Martínez-Mora, G. Larosa, S. Adrián-Martínez**

Universitat Politècnica de València representing the KM3NeT Consortium,
C/ Paranimf 1, E-46730 Gandia, Spain

E-mail: cdavid@upv.es (C.D. Llorens), mardid@fis.upv.es (M. Ardid)

**ABSTRACT:** We describe the sound emission board proposed for installation in the acoustic positioning system of the future KM3NeT underwater neutrino telescope. The KM3NeT European consortium aims to build a multi-cubic kilometre underwater neutrino telescope in the deep Mediterranean Sea. In this kind of telescope the mechanical structures holding the optical sensors, which detect the Cherenkov radiation produced by muons emanating from neutrino interactions, are not completely rigid and can move up to dozens of meters in undersea currents. Knowledge of the position of the optical sensors to an accuracy of about 10 cm is needed for adequate muon track reconstruction. A positioning system based on the acoustic triangulation of sound transit time differences between fixed seabed emitters and receiving hydrophones attached to the kilometre-scale vertical flexible structures carrying the optical sensors is being developed. In this paper, we describe the sound emission board developed in the framework of KM3NeT project, which is totally adapted to the chosen FFR SX30 ultrasonic transducer and fulfils the requirements imposed by the collaboration in terms of cost, high reliability, low power consumption, high acoustic emission power for short signals, low intrinsic noise and capacity to use arbitrary signals in emission mode.

---

[*] Corresponding author.

**Contents**

**1. Introduction**

The Sound Emission Board (SEB) presented in this article is part of the long baseline acoustic positioning system proposed for the future underwater neutrino telescope KM3NeT that will be located at the Mediterranean Sea.

The KM3NeT Consortium [1] aims to build an underwater neutrino telescope of at least one cubic kilometre volume. Due to the large instrumented volume needed, many of the hardware solutions adopted in the first undersea neutrino telescope ANTARES [2], which is taking data since 2008 with an effective area of $0.1 km^2$, cannot directly be applied to KM3NeT; the costs and production period would be prohibitive. New designs of the detector sub-systems are required.

The detection principle used in underwater neutrino telescopes is based on the detection of the Cherenkov light produced by muons coming from $\nu_\mu$ interactions with matter in or near to the telescope. Both ANTARES and the future KM3NeT require knowledge of optical module positions with an accuracy of about 10 cm in order to properly reconstruct muon tracks detected by the photomultipliers. Since the mechanical structures holding the optical sensors are not completely rigid and can move due to sea currents, a positioning system is mandatory. In ANTARES, the positioning calibration system provides the positions of the optical modules with accuracy better than 10 cm [3]. While KM3NeT will be 20 times larger, the ANTARES positioning calibration system will not be directly scalable: a new design of the system has been necessary. The Acoustic Positioning System (APS) included in the KM3NeT general positioning calibration system is a Long Baseline System composed of a set of acoustic transceivers and the associated electronics (the subject of this paper) and an array of acoustic receivers (hydrophones) rigidly attached to the telescope mechanical structures. This APS should provide the position of the telescope mechanical structures, in a geo-referenced coordinate system with accuracy better than 1m (for a good pointing accuracy of the telescope) and also the positions of the 10000 + optical modules during continuous telescope operation in varying deep sea currents with a precision of about 10 cm. In addition, the acoustic devices installed in KM3NeT will be used in studies related to sea sciences (e.g. bioacoustics, geophysics, etc.) and possibly in the acoustic detection of ultra high energy neutrinos.

An important aspect of the APS system is the transceiver. Following the idea of reducing cost and increasing reliability, a new design for this system has been proposed, and a prototype has been developed [4,5]. It basically consists of two parts: the acoustic transducer and the electronics: the Sound Emission Board (SEB). The selected transducer for this system is the FFR SX30 transducer manufactured by Sensortech. The most important reasons for choosing this transducer are:

- it can provide a reasonable acoustic power level for long distance transmission. Considering the transmission power level given by the manufacturer (133 dB Ref. 1 µPa/Volt @ 1 m) and the sensitivity of the receiver hydrophones developed by INFN – LNS [6], the required transducer excitation voltage has been calculated (figure 1). It has been found that 500 V is enough for reasonable levels in the receivers for distances up to 1.9 km. The calculation has been made for the 30 kHz resonant frequency: the impedance at this frequency is (130-1000j) $\Omega$;
- it has an 'unlimited' depth (pressure) of operation according to the manufacturer specifications. Transducers were tested in a hyperbaric tank up to 440 bars checking the acoustic sensitivity for different frequencies under high pressure. Results obtained for this study were clearly satisfactory and no significant pressure-dependent changes in the expected behaviour of the transducer [4] were observed;
- the operation frequency range is 20kHz – 40 kHz; the optimum frequency range for positioning purposes.

## 2. The Sound Emission Board.

In this section we describe the specifications required for SEB operation in KM3NeT, its different parts, and the different solutions adopted to meet the imposed specifications, as follows;
- power consumption less than 1W at 5V (control part) & 12V (power part);
- low speed communication port required to configure the board and load arbitrary signals from shore. RS232 or RS485 communication ports have been implemented;
- time synchronization must work with an accuracy of about 1µs;
- the SEB must be able to work in emission and reception mode;
- trigger used in emission mode will be through a LVDS signal;
- the entire system life expectancy must exceed twenty years.

### 2.1 Basic block diagram.

In figure 2 the basic block diagram of the SEB is shown. The transducer is shown at the top of the diagram. The switching block allows transducer connection to the SEB driver or to an external acoustic board [6], which digitises the transducer signal when it is operated in receiver mode. When the transducer is operated as emitter, it is connected to the power amplifier through an impedance matching network. As the instantaneous power available through the telescope 12V DC distribution system is less than that needed to excite the transducer to cover long distances, it has been necessary to implement an energy storage block. In the lower part of the block diagram the signal generator that drives the power amplifier is shown. It has two inputs, one for the low bitrate communication port and one for the trigger signal.

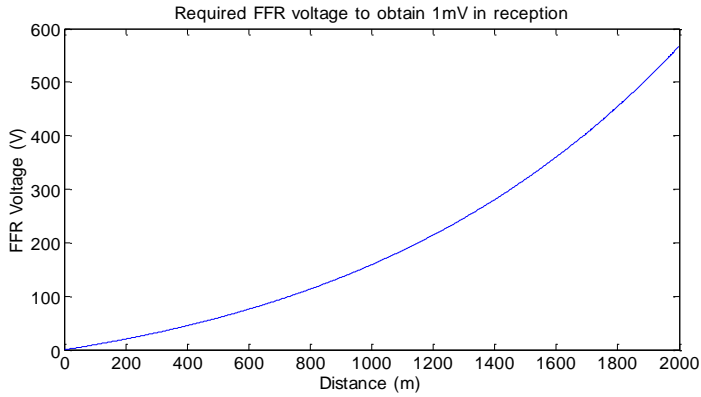

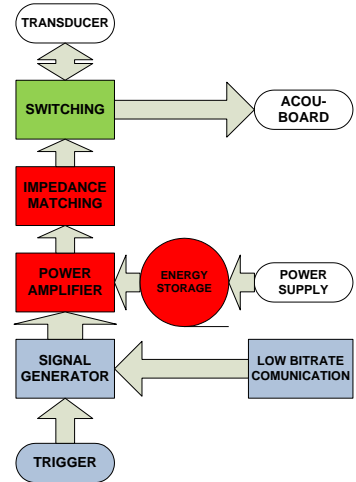

Fig.1. Required transducer excitation voltage needed to obtain the minimum signal a function of reception distance.

Fig.2. Conceptual block diagram of the SEB.

**2.2 Impedance matching block.**

As shown previously, the generation of an acoustic signal with enough power to be detectable at long distances requires more than 12VDC offered to feed the SEB. For this reason, a transformer - that also plays the main role of impedance matching - has been implemented in the SEB. While standard impedance matching networks interpose several inductors and capacitors between the power amplifier and transducer, the typical tolerances of inductors and capacitors (respectively 20% and 10%) would cause unacceptable variations in latency time between different SEB boards. For this reason in the SEB we only use a single transformer for adapting the impedance, as shown in figure 3. Two variants of transformer have been implemented in the SEB prototype; the first with 1:20 turns ratio ($480V_{pp}$ output, without load), and the second with 1:30 turns ratio ($720V_{pp}$ output, without load).

**2.3 Energy storage blocks.**

In our system, we use capacitors to store the required transmission energy. The solution allows for fast charging, and correspondingly short time delays between successive emissions (the usual mode of operation is a high-power emission of a few ms duration every few seconds). The solution also offers a long life expectancy. The minimum capacitance needed for the emission has been calculated. Using the 1:30 transformer, the maximum output voltage achieved is 720 $V_{pp}$; with this voltage and the impedance of the transducer, the power applied is 65 $W_{rms}$. As the efficiency of the transformer is more than 80% and the power amplifier has more than 90% efficiency, the power required to feed the capacitor is around $90W_{rms}$ (7.5A @ 12V).

Considering the maximum length of the signal of 1ms and a maximum capacitor discharge of 1V, the required capacitance exceeds 7.5mF.

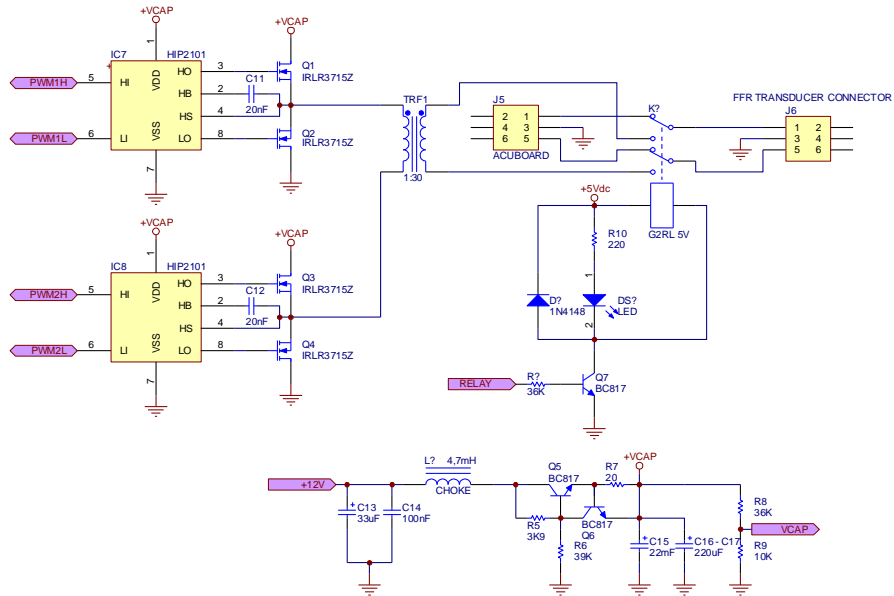

Fig.3. Power part schematic

**2.3.1 The aluminium capacitor**
For the SEB first prototype a 22mF@16V capacitor from Panasonic was selected. The ESR (Equivalent Series Resistance) of this capacitor is 32 mΩ at 20 kHz. Two different models were selected, the ECOS1CP223CA with 3000 hours of life at 85ºC and the ECOS1AA223BA with 2000 hours of life at 105 ºC. Using the manufacturer's lifetime/temperature projection equation the life expectancies at 25ºC obtained - considering the worst case (15ºC of ripple current temperature rise) - for the ECOS1CP223CA and ECOS1AA223BA were 7.7 years and 20.6 years, respectively. These values are given with a 60% confidence level.

**2.3.2 The tantalum capacitor.**
During tests of the first prototype a maximum transducer excitation of 400$V_{pp}$ was seen using the 30:1 transformer (the value without the transducer was close to 720$V_{pp}$). Some losses in the power amplifier and energy storage blocks were also observed. In order to correct this in the second prototype, the use of one hundred parallel tantalum capacitors has been considered. The individual tantalum capacitors usually have a much better ESL (Equivalent series inductance) than the aluminium capacitors, but they each have lower capacitance and slightly more ESR. However, using one hundred capacitors offers the possibility of multiplying the capacitance and reducing the ESL and ESR by one hundred. The MIL-HDBK-217F[1] standard was used to determine the reliability of a single capacitor. Calculations were made for two different VISHAY models; the first of these (TR3E227K016C0100) is less expensive, but the second (TR3E227K016C0100) has much greater reliability due to the bigger difference between the rated voltage and the used voltage (12 V). Table 1 presents the main parameters for each type.

---

[1] MIL-HDBK-217F, Reliability Prediction of Electronic Equipment, is a military standard that provides failure rate data for many military electronic components.

Table1. Characteristics (Rated Voltage, ESR and Mean Time Between Failures) for tantalum capacitors and for the 100 parallel configuration.

| MODEL | Rated V | ESR | ESR 100 parallel | MTBF years single | MTBF years 100 parallel |
|---|---|---|---|---|---|
| TR3E227K016C0100 | 16 | 0.1 Ω | 1mΩ | 2200 | 22 |
| 597D227X9020R2T | 20 | 0.08 Ω | 0.8 mΩ | 50000 | 500 |

### 2.4 Power amplifier block.

The power amplifier solution adopted is a class D amplifier formed by a full bridge. The full bridge is composed of an Intersil HIP2101 MOSFET driver[2] and four IRLS3715Z MOSFETs from International Rectifier. The main characteristics for the MOSFETs are the low $R_{on}$ (transistor ON resistance) of 11 mΩ, the fast switching (13ns rise and 4.7ns fall) and low gate charge ($Q_G = 7.2nC$). Other important characteristics are the high drain current ($I_d = 200$ A, 0.1 ms signal) and the maximum drain to source voltage (20V). The SEB power part contains the transducer switching relay, the MOSFETs and their drivers, the energy storage capacitor with the corresponding current limiter and filter for the charging process.

For the new prototype we are considering several Infineon MOSFETs (all with $R_{on} < 2m\Omega$). The most interesting types are not yet on the market. But from tests so far made our choice is the model BSC020N03LSG ($R_{on}=1.7m\Omega@V_{GS}=10V$, $Q_c=15nC$, 7ns switching time and $I_d=400A@0.1ms$ signal). Considering the lowest capacitor ESR and MOSFET $R_{on}$ of the new prototype, we are improving the power part $Z_{out}$ from 54mΩ to 5mΩ and - since the transducer impedance in the primary of the 30:1 transformer is close to $(0,14-1j)\Omega$ - the efficiency of the power part will be very much improved.

### 2.5 Signal generator block

Having chosen the full bridge power amplifier, it must be fed with squared signals. This is not a problem if we want to send squared signals like an MLS (Maximum Length Sequence), which is a very useful signal that is extensively used in electro-acoustic measurements. The main characteristics of this signal are the plain spectrum and the non-correlation with any other signal. It can be used to obtain the impulse response of the entire system and for time of flight measurements. Moreover, if we wish to send standard sinusoidal or arbitrary signals we can take advantage of the fact that the transducer and the transformer are good band pass filters in our band of interest: we can emit a square signal in the band and all the highest frequencies that are out of the working band will be removed. We can also use square signals and obtain sinusoidal signals in the emission, although the best technique - should we wish to send arbitrary signals by generating squared signals - is using PWM (Pulse Width Modulation) with a modulation frequency outside the main band. To implement PWM we must vary the width of the square signal in direct relation to the voltage of the desired signal (0-100% Pulse width). The classic way to do this is to compare the desired signal with a triangular or sawtooth signal in order to obtain a square signal at the output of the comparator. After the amplification, the desired signal is integrated (filtered by the transformer and the transducer) and the median value of the square signal is obtained. This median value is the desired signal.

---

[2] We are considering the HIP2111 for the second prototype due to better behavior at higher current.

For the signal generator we have decided to use the Microchip "Motor Control" function inside most of the DSPic microcontroller series. For the first prototype we selected the DSPic33FJ256MC710. This microcontroller has 40 MIPS of processing power, signal processing specific instructions, enough FLASH and RAM for our purposes and a 10bits@1MSps ADC converter. The "Motor Control" function is basically a digital counter that works with the main frequency of the microcontroller (40MHz). This device has all the necessary components to work with full MOSFET bridges (symmetric outputs, dead time generators, etc.), and for this reason matches perfectly for our purposes. We use two of the function modes of the motor control. The first is the "free run" mode; in this mode we can obtain a pulse with modulation similar to the classic one that compares the modulation signal with a saw tooth. Using this mode when the counter arrives to the PxTPER[3] (maximum) value it is reset and starts again from zero. The device has also other comparators in order to establish the width of the pulse. The second mode we use is the "up/down" mode. The only difference between this and the previous mode is that when the counter arrives to PxTPER it starts counting down instead of resetting. This mode is similar to the classic PWM modulation that compares the signal with a triangular wave. With this mode we obtain a more symmetric square signal with fewer harmonics.

**2.6 The Firmware**

We have developed different firmware versions in order to adapt our board to the communication prerequisites of the different test installations at the ANTARES and NEMO neutrino telescope sites (in ANTARES: with MODBUS[4] over RS485; in NEMO: console over RS232). However the basic working process is described in figure 4. The basic firmware has three main parts. The first is the processing of commands that arrive at the low bitrate communication port: these commands usually are for configuring the board or defining the signal to be emitted. The second block is in the main part of the program and is aware of the trigger port in order to start the emission when a trigger signal arrives. The third block is an interrupt code that works when the "Motor Control" counter arrives to the PxTPER register. This code changes the registers of the "Motor Control" device in order to obtain the next cycle of the desired square signal (frequency, pulse width, etc.)

---

[3] PxTPER is the PWM time base period register. It sets the maximum value of the counter.
[4] MODBUS is a communication protocol used in industrial environments.

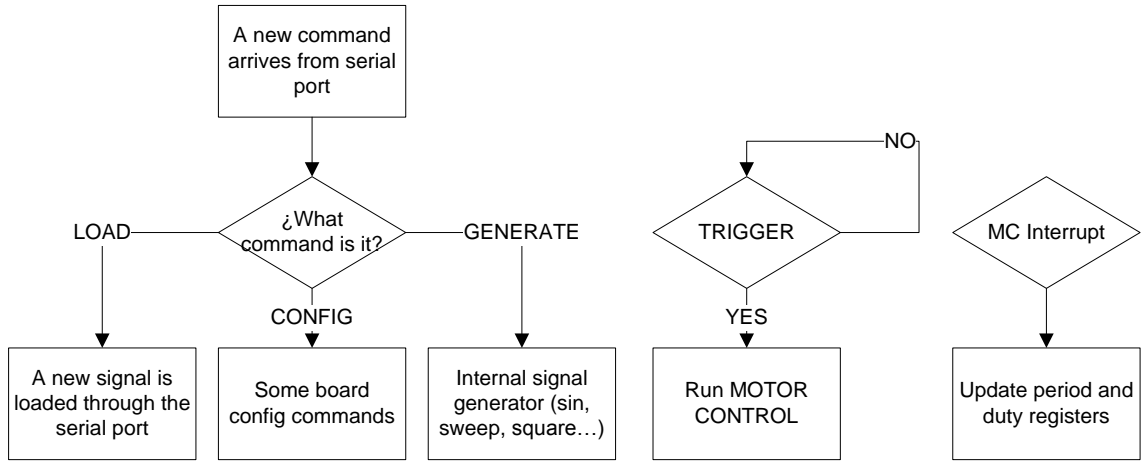

Fig. 4: Diagram of the firmware working process.

## 3. Tests

Laboratory tests have verified that the different components behave as expected. The whole prototype has been fully tested in the lab in order to check that the power consumption, acoustic power emission, time stability, communication and configuration capability, and reliability conformto the specifications. During this process, several changes in the components were made to improve the board (as an example, a more powerful micro-controller is being used with respect to previous versions [4]). The final system described here has been shown in the lab tests to fulfill all the requirements and specifications: power consumption less than 1 W, good timing precision (measured latency around 5 μs with a stability better than 1 μs), well adapted to the standards used in deep-sea neutrino telescopes [7]. However, as mentioned, minor modifications have been derived and will be included in future versions in order to increase the efficiency of the power amplifier block.

In order to test the SEB prototypes under real conditions (in the hostile deep-sea environment) and to test their compatibility with neutrino telescope infrastructures in general and the positioning system in particular, our system is being integrated in two different sites: the Instrumentation Line of ANTARES (2500m depth: off Toulon, France) and the NEMO phase II[8] (3600m depth: off Capo Passero, Italy). The tests performed during the integration have shown the compatibility of the SEB with the rest of the elements of these neutrino telescopes, and we are presently waiting for the final deployment and deep sea connection of these KM3NeT prototypes to be able to test the SEB in situ.

## 4. Conclusions

We have presented and described the solutions adopted for the Sound Emission Board for the KM3NeT acoustic positioning system. Our system is considered for the implementation in the final configuration of the KM3NeT detector. The first prototype developed has been tested in this framework. In order to test the system under real conditions (deep-sea), our system has been integrated in the Instrumentation Line of ANTARES (deployed and waiting for connection), and is also being integrated in NEMO phase II. Presently, a new prototype of the SEB improving the output impedance of the power part is being developed in order to be able to feed the transducer more efficiently.

## Acknowledgments

This work was supported by the European Commission through the KM3NeT Design Study (FP6, contract no. DS 011937) and Preparatory Phase (FP7, grant no. 212525) and also by the Ministerio de Ciencia e Innovación (Spanish Government), project references FPA2009-13983-C02-02, ACI2009-1067, Consolider-Ingenio Multidark (CSD2009-00064). It was also funded by Generalitat Valenciana, Prometeo/2009/26.